\documentclass[12pt,a4paper]{article}
\usepackage{graphicx}
\usepackage[english]{babel}
\usepackage{amsmath}
\usepackage{amssymb}
\usepackage{amsfonts}
\numberwithin{equation}{section}

\begin{document}
\def\pp{{\, \mid \hskip -1.5mm =}}
\def\cL{{\cal L}} 
\def\beq{\begin{equation}}
\def\eneq{\end{equation}}
\def\bea{\begin{eqnarray}}
\def\enea{\end{eqnarray}}
\def\tr{{\rm tr}\, }
\def\nn{\nonumber \\}
\def\e{{\rm e}}

\title{\textbf{Kaluza-Klein reduction of relativistic fluids and their gravity duals}}
\author{Adriana Di Dato}
\date{}
\maketitle

\vskip 10pt
\begin{center}
\textit{Departament de F\'\i sica Fonamental and\\ Institut de Ci\`encies del Cosmos, Universitat de Barcelona,\\ Mart\'\i\  i Franqu\`es 1, ES-08028, Barcelona, Spain.}\\

\vskip 10pt
{\small E-mail: adidato@ffn.ub.es}
\end{center}

\vspace{25pt}

\begin{abstract}
We study the hydrodynamics of relativistic fluids with several conserved global charges (i.e., several species of particles) by performing a Kaluza-Klein dimensional reduction of a neutral fluid on a $N$-torus. Via fluid/gravity correspondence, this allows us to describe the long-wavelength dynamics of black branes with several Kaluza-Klein charges. We obtain the equation of state and transport coefficients of the charged fluid directly from those of the higher-dimensional neutral fluid. We specialize these results for the fluids dual to Kaluza-Klein black branes.
\end{abstract}
%%%%%%%%%%%%%
\newpage

\tableofcontents

\section{Introduction}
\label{1}

Kaluza-Klein dimensional reduction is a well known method to obtain solutions to a gravitational theory coupled to a Maxwell field, plus a scalar (dilaton) field. Velocities (or momenta) along the compactified direction result in electric charges in the reduced theory \cite{Pope}. Thus, if we take a neutral black string solution of the vacuum Einstein theory, perform a boost along the direction of the string and then dimensionally reduce in this direction, we obtain an electrically charged black hole of the Einstein-Maxwell-dilaton theory, for a particular value of the dilaton coupling \cite{Horowitz}. 

It should be clear that this method is not exclusive to gravitational theories. The identification between momenta along the internal direction and conserved charges in the reduced theory is in fact generic. Note, however, that in a non-gravitational theory, one obtains charges of a global symmetry group --- e.g., a global $U(1)$ for reduction in a circle --- while in the gravitational case, since the relevant spacetime symmetries are gauged, they are charges of a gauge symmetry group.

In this article we are interested in applying the Kaluza-Klein procedure to relativistic hydrodynamics. That is, we begin with a relativistic fluid without any conserved particle number in $p$ spatial dimensions, where $p-N$ of these are non-compact directions and $N$ of them form an $N$-torus. We assume that none of the fluid variables depend on the internal directions, but the fluid can have non-trivial velocity along them. These velocities give internal momenta that in the reduced theory appear as conserved global charges, i.e., particle numbers for $N$ different species. For a perfect fluid, this reduction is a straightforward one. Of more interest is the reduction of the first-order dissipative terms. Viscosity of the higher-dimensional fluid in the internal directions gives rise not only to viscosities but also conductivities in the reduced theory.

We shall do our analysis for a generic relativistic fluid in $p$ spatial dimensions with no conserved particle number, without assuming any specific equation of state nor constituent relation for its first-order transport coefficients. When applying our results to particular fluids, we will consider a class of recent interest in the context of dual relations between fluid dynamics and black brane dynamics. These are the fluids that correspond to neutral black $p$-branes of the vacuum Einstein theory, and which feature in the blackfold approach  to black brane dynamics \cite{Emparan,Joan}. Ref.~\cite{Joan} developed the dictionary between the spacetime fluctuations of these black branes and the fluctuations of specific fluids. Using this mapping, our results yield a mapping between the dynamics of charged black branes in Kaluza-Klein theory and the hydrodynamics of certain charged fluids. The map includes their fluid equation of state and first-order transport coefficients. Note that the Kaluza-Klein black brane solutions differ from other charged black branes in their coupling to the dilaton. While the dilaton plays no direct role in the dual fluid description, since it is not associated to any conserved quantity of the black brane, the value of its coupling affects the equation of state and constituent relations.

There have been some previous studies of Kaluza-Klein reduction in the context of fluid/gravity correspondences \cite{Blaise,Blaise2,Milena,Kanitscheider,Kanitscheider2}. However, these have been restricted to the fluids that are dual to AdS black branes, and moreover they only work out explicitly the cases of circle \cite{Blaise} and 2-torus reduction \cite{Milena}. The results of \cite{Blaise,Milena} can be mapped, via the AdS/Ricci-flat connection of \cite{Joan2}, to our results for circle reductions of a neutral vacuum black brane. 
Vice versa, our results can be readily translated into results for AdS black branes with $N$ different charges using this mapping.
The perfect fluid dynamics of asymptotically flat charged branes (with
arbitrary dilaton coupling) was studied in \cite{Caldarelli,Emparan:2011}.
Dissipative effects of non-dilatonic asymptotically flat charged branes have
been analyzed in \cite{Gath}. The first-order hydrodynamics of asymptotically
flat black D3-branes has been studied in \cite{Emparan:2013}, but the charge in this
case can not be redistributed along the worldvolume and therefore the
dynamics is qualitatively different. Moreover, in \cite{Armas, Armas2} the Kaluza-Klein approach has
been applied, in a slightly different manner, 
to obtain
first-derivative corrections of charged black brane (extrinsic) dynamics.

In our opinion it is useful to treat the Kaluza-Klein reduction of fluids separately from any specific fluid/gravity dualities. First, this makes clear how the procedure stands on its own within the context of hydrodynamics without any reference to General Relativity. Second, by not tying the reduction to any particular fluid, we achieve a large degree of generality. Clearly, the method can be extended to the case in which the higher-dimensional fluid carries a particle number or some other property, but we will not pursue this in the present article.

%%%%%%%%%%%%%%%%%%5

\section{Hydrodynamic Kaluza-Klein ansatz and \\reduction of the perfect fluid}
Let us consider a neutral relativistic fluid in flat space-time in $p+1$ dimensions.
The hydrodynamical behaviour of this fluid is governed by the stress energy tensor conservation equations $\partial_A T^{AB}=0$.
In general we split the stress energy tensor into a perfect fluid and a dissipative part,
\bea
T_{AB}=T_{AB}^{pf}+T_{AB}^{diss} \label{eqn:generT}\,.
\enea 
The perfect fluid part is given in terms of
the energy density $\epsilon$, the pressure $P$ and the normalized velocity field $u^A$ by
\beq\label{pf}
T_{AB}^{pf}=(\epsilon +P)  u_A  u_B + P g_{AB}
\eneq
while, to first derivative order, the dissipative part is
\beq\label{eqn:diss}
T_{AB}^{diss}=-2\eta \sigma_{AB}-\zeta P_{AB} \theta
\eneq
where  $\eta$ and $\zeta$ are the shear and bulk viscosities and $\theta$, $ \sigma_{AB}$ and $P_{AB}$ are defined as
\bea
\theta&=&\partial_A u^A\label{espa},\\
\sigma_{AB}&=& P_A^C P_B^D\partial_{(C} u_{D)}-\frac{1}{p}\theta P_{AB}\label{shear}, \\
P_{AB}&=&g_{AB}+u_A u_B\label{orto}\,.
\enea
A complete description of the fluid requires the specification of  the equation of state, namely the relation between $P$ and $\epsilon$, and of the viscosities. For the most part we will keep them general, and will only specify them in sec.~\ref{6}.
Furthermore, we naturally assume that this uncharged fluid is described in the Landau frame where $u^A  T_{AB}^{diss}=0$.

We assume that the spacetime in which the fluid moves contains $N$ compact directions that form an
$N$-torus
\begin{equation}\label{ansatz}
d\hat s^2= \sum_{j=1}^{N} dy_j^2 +\eta_{ab} d\sigma^a d\sigma^b \,,
\end{equation}
where the metric $\eta_{ab}$ is the Minkowski metric in $p-N+1$ spacetime dimensions and the coordinates $y_j$ are identified with periodicity $2\pi R_j$. 
We take the fluid to move with non-zero velocity along the $N$  compactified dimensions. On Kaluza-Klein reduction this will give rise to charges in the reduced fluid.

The Kaluza-Klein ansatz for the field assumes that none of the fluid variables depend on the internal directions $y_j$.
The velocity profile is
\bea\label{boost}
u_a=\hat{u}_a\prod_{i=1}^{N} \cosh\alpha_i,\qquad
u_{y_j}= \sinh\alpha_j \prod_{k=1}^{j-1} \cosh\alpha_k \,,\quad j=1,\dots, N
\enea
where $\alpha_i$ are boost parameters characterizing the velocity along the compact directions and $\hat{u}_a$ is the velocity in the reduced spacetime, which is unit-normalized with respect to $\eta_{ab}$,
\beq
\hat{u}^a\hat{u}^b \eta_{ab}=-1\,.
\eneq
Note that it should be possible to formulate an ansatz for the velocity field where the different boosts enter in a manner that preserves the local symmetry $SO(N)$ that rotates them (this is broken globally by the compact size of the torus). The above ansatz does not show this, but it is a convenient one for our calculations.\footnote{Our choice is in this sense analogous to choosing polar coordinates for a sphere, which allows easy explicit calculation but obscures the rotational symmetry.}

Let us apply this Kaluza-Klein reduction ansatz to the perfect fluid stress energy tensor.
Substituting \eqref{boost} in \eqref{pf} we obtain 
\bea\label{eqn:perfectgen}
T_{ab}^{pf}&=&V  [(\epsilon + P)  \hat u_a \hat u_b \prod_{i=1}^{N} \cosh^2\alpha_i  + P\eta_{ab}]\label{eqn:perfectgen1}\,,\\
T_{ay_j}^{pf}&=&V (\epsilon + P)\sinh\alpha_j\hat u_a\prod_{i=1}^{N} \cosh\alpha_i \prod_{k=1}^{j-1} \cosh\alpha_k  \label{eqn:perfectgen2}\,,\\
T_{y_j y_{j'}}^{pf}&=&V  [(\epsilon + P)  \sinh\alpha_{j'} \sinh\alpha_j\prod_{k=1}^{j-1} \cosh\alpha_k\prod_{i=1}^{j'-1} \cosh\alpha_i  +\eta_{jj'}P]\label{eqn:perfectgen3}\,,\qquad
\enea 
where $V=\prod_{j=1}^{N} (2\pi R_j)$ is volume of the torus.
The factor $V$ appears because $T_{AB}$ refers to densities so we have to include the internal volume  we are going to integrate out.
The form of the stress energy tensor in the reduced theory is
\bea
\hat T_{ab}^{pf}&=&(\hat\epsilon +\hat P) \hat u_a \hat u_b +\hat P \eta_{ab}\label{eqn:perfectgenbb1}\,,\\
\hat T_{ay_j}^{pf}&=& u_a \hat q_j\label{eqn:perfectgenbb2}\,,
\enea
where
\bea
\hat P &=&P  V\label{pres}\,,\\
\hat\epsilon &=&\hat P (-1+\prod_{i=1}^{N} \cosh^2\alpha_i)+ \epsilon V \prod_{m=1}^{N} \cosh^2\alpha_{m}\label{en}\,,\\
\hat q_j&=&(\hat P+\hat \epsilon) \frac{\sinh\alpha_j}{ \prod_{i=j}^{N} \cosh\alpha_i}\label{car}\,.   
\enea
This yields not only the energy density and pressure in the reduced theory, but also a set of $N$ charge densities $\hat q_j$, one for each boost parameter $\alpha_j$.

The temperature of the reduced fluid is given by 
\beq\label{tempe}
\hat T=\frac{T}{ \prod_{i=1}^{N} \cosh\alpha_i}
\eneq
due to the fact that we  have changed the timelike Killing vector.
From the conservation of the entropy current for the initial fluid, we can read off the reduced entropy density given by
\beq\label{redentropy}
\hat s=s  V \prod_{i=1}^{N}\cosh\alpha_i\,.
\eneq
From the Euler relation  
\beq
\hat P+\hat\epsilon = \hat T \hat s+\sum_{j=1}^N\hat q_j \hat\mu_j
\eneq
the chemical potential for each charge takes the form
\beq\label{mu}
\hat\mu_j= \frac{\sinh\alpha_j }{\prod_{i=j}^{N} \cosh\alpha_i}\,.
\eneq

Since we assume that the first law is satisfied for the initial neutral fluid, it is possible to verify that the same is true for the reduced fluid. 
The neutral fluid obeys the law
\beq
d \epsilon= T d S
\eneq
from which it follows
\beq
d \hat\epsilon= \hat T d \hat S+ \sum_{i=1}^N \hat\mu_i d\hat q_i
\eneq
using 
\bea
\hat \epsilon+ \hat P&=&V(\epsilon+  P)\prod_{i=1}^{N} \cosh^2\alpha_i
 \qquad\mbox{from Eq.\eqref{en}},\\
d\alpha_{j+1}&=&d\alpha_{j} \frac{\tanh\alpha_{j+1}}{\sinh\alpha_j\cosh\alpha_j}\qquad\mbox{from Eq.\eqref{eqn:relazvarjdif}}
\enea
and that $\epsilon+ P=TS$.

\section{Reduction of dissipative terms}
The Kaluza-Klein reduction has given us a charged fluid. When including dissipative terms we expect the presence of another set of transport coefficients, namely a heat conductivity matrix. 
These coefficients measure the response of the charge current to changes in temperature and in chemical potential.

In order to reduce the first-derivative terms in the stress energy tensor, we need to express the expansion $\theta$ defined in Eq.(\ref{espa}) in terms of the new velocities. We find that
\beq\label{eqn:thetagen}
\theta= \prod_{i=1}^{N}\cosh\alpha_i(\hat \theta+\sum_{k=1}^{N} \tanh\alpha_k\hat u_a\partial^a \alpha_k)\,,
\eneq
where $\hat\theta=\partial^a \hat u_a$.

The equation of conservation of the stress energy tensor relates the gradients of the rapidities to $\hat \theta$ as
\beq
\hat u_a \partial^a \alpha_j =\hat\theta\frac{\cosh\alpha_{j}\sinh\alpha_{j}\prod_{l=j+1}^{N} \cosh^2\alpha_l}{1+(-1+\epsilon')\prod_{i=1}^{N} \cosh^2\alpha_i}\,,
\eneq
where 
\beq
\epsilon'=c_s^{-2}=\frac{\partial \epsilon}{\partial P}\,,
\eneq
where $c_s$ is the speed of sound.
The explicit calculation can be found in the Appendix \ref{eqmotion}.
Substituting this result in Eq.\eqref{eqn:thetagen} the expansion becomes
\beq\label{eqn:thetagen1}
\theta=\hat \theta \frac{\epsilon'\prod_{i=1}^{N}\cosh^3\alpha_i}{1+(-1+\epsilon')\prod_{l=1}^{N}\cosh^2\alpha_l}\,.
\eneq
The orthogonal projectors tensor in Eq.\eqref{orto} is given by
\bea
P_{ab}&=&\prod_{i=1}^{N}\cosh^2\alpha_i  \hat u_a \hat u_b + \eta_{ab}\,,\\
P_{ay_j}&=& \hat u_a\sinh\alpha_j  \prod_{i=1}^{N} \cosh\alpha_i \prod_{k=1}^{j-1} \cosh\alpha_k \,, \\
P_{y_j y_{j'}}&=& \sinh\alpha_{j'} \sinh\alpha_j\prod_{k=1}^{j-1} \cosh\alpha_k \prod_{i=1}^{j'-1} \cosh\alpha_i +\eta_{jj'}\,.
\enea
The shear viscosity tensor in Eq.(\ref{shear}) takes the form
\bea
\sigma_{ab}&=&P^C_a P^D_b\partial_{(C} u_{D)}-\frac{ P_{ab}\theta}{p}\nonumber\\
&=&\sum_{i=1}^N P^c_{(a} P^{y_i}_{b)}\partial_c u_{y_i}+P^c_a P^d_b\partial_{(c} u_{d)}-\frac{ P_{ab}\theta}{p}\label{eqn:ab}\,,
\enea
\bea
\sigma_{ay_j}&=&P^C_a P^D_{y_j}\partial_{(C} u_{D)}-\frac{ P_{a{y_j}}\theta}{p}\nonumber\\
&=&P^c_a P_{y_j}^d\partial_{(c} u_{d)}+\sum_{i=1}^N P^{y_i}_{(a} P_{y_{j})}^d\partial_d u_{y_i}
-\frac{ P_{ay_j}\theta}{p}\label{eqn:ay}\,,
\enea
\bea
\sigma_{y_j y_{j'}}&=&P^C_{y_j} P^D_{y_{j'}}\partial_{(C}u_{D)}-\frac{ P_{y_j y_{j'}}\theta}{p}\nonumber\\
&=&P^c_{y_j} P^d_{y_{j'}}\partial_{(c} u_{d)}
+\sum_{i=1}^N P^{y_i}_{(y_j} P^d_{y_{j'})}\partial_{d} u_{{y_i}}-\frac{ P_{y_j y_{j'}}\theta}{p}\label{eqn:yy}\,.
\enea
We have all the ingredients to compute the new transport coefficients. However, our reduced fluid is not in the Landau frame. Indeed, we find that $u^A T_{AB}=0$ implies
\bea
u^a T_{ab}^{diss}&+&\sum_{j=1}^{N} u^{y_j}T_{{y_j}b}^{diss}=0\,,\\
u^a T_{a y_j}^{diss}&+&\sum_{j'=1}^{N} u^{y_{j'}}T_{{y_{j'}} y_j}^{diss} =0\,,
\enea
or using Eq.(\ref{boost})
\bea
\hat u^a T_{ab}^{diss}&=&-\frac{\sum_{j=1}^{N}\sinh\alpha_j}{\prod_{i=j}^{N} \cosh\alpha_i} T_{{y_j}b}^{diss}\label{eqn:Laundau1}\,,\\
\hat u^a T_{a y_j}^{diss}&=&-\frac{\sum_{j'=1}^{N}\sinh\alpha_{j'}}{\prod_{i=j'}^{N} \cosh\alpha_i}T_{{y_{j'}} y_j}^{diss}\label{eqn:Laundau2}\,.
\enea
This means that we cannot directly extract the coefficients from the reduced stress energy tensor but we need to introduce some frame-invariant formulae. 
In \cite{Bhattacharya2} was proposed an efficient way to extract those coefficients based on a general dissipative correction to the stress energy tensor and the charge currents.  In order to avoid unphysical solutions we require the semi-positivity of the divergence of local entropy current.
Following the same procedure as in \cite{Bhattacharya2} and generalizing the result for $N$ charges we construct frame invariant formulae
\bea
\hat P^a_c \hat P^b_d T_{ab}^{diss} - \frac{1}{p-N} \hat P_{cd} \hat P^{ab} T_{ab}^{diss}=-2 \hat\eta \hat\sigma_{cd}\label{eqn:etagen}\,,\\
\hat P^b_a\Big( T_{by_j}^{diss} +\frac{\hat q_j}{\hat\epsilon+\hat P} \hat u^c T^{diss}_{cb}\Big) =-\sum_{j'=1}^{N}\hat\kappa_{jj'} \hat P^b_a\partial_b\left(\frac{\hat\mu_{j'}}{\hat T}\right)\label{eqn:kappagen}\,,\qquad\\
\frac{\hat P^{ab} T_{ab}^{diss}}{p-N}-\frac{\partial \hat P}{\partial \hat\epsilon}\hat u^a \hat u^b T_{ab}^{diss}+\sum_{j=1}^{N}\frac{\partial \hat P}{\partial \hat q_j}\hat u^a T_{ay_j}^{diss} =-\hat\zeta \hat \theta\label{eqn:zetagen}\,.
\enea
Using these we can extract the viscosities $\hat\eta$, $\hat\zeta$ and the matrix of conductivities $\hat\kappa_{jj'}$. The derivative $\partial \hat P/ \partial \hat\epsilon$ is evaluated at constant charges, while $\partial \hat P/\partial  \hat q_j$ are evaluated keeping fixed the energy density and the other charges $q_{k\neq j}$.\\
We obtain
\bea
\hat\eta=\eta V \prod_{i=1}^{N} \cosh\alpha_i\label{eqn:etaflui}\,,
\enea
\bea
\hat\kappa_{jj}&=& \eta V\hat T \Big{(}1-\frac{\sinh^2\alpha_j}{\prod_{l=j}^{N} \cosh^2\alpha_l}\Big{)}\prod_{i=1}^{N} \cosh\alpha_i\label{eqn:kappafluidjj}\,,\\
\hat\kappa_{jk}&=&\hat\kappa_{kj}=-\eta V\hat T\frac{\sinh\alpha_j\sinh\alpha_{k}}{\prod_{i=j}^{N} \cosh\alpha_i}\prod_{l=1}^{k-1} \cosh\alpha_l\label{eqn:kappafluidjk}\,,\quad \mbox{with}\,\, k\neq j\,,
\enea
\bea
\hat\zeta&=&2\eta V\prod_{i=1}^{N} \cosh\alpha_i\Big{[} \frac{1}{p-N}\label{eqn:zetafluid}\\
&+&\frac{(-1+\prod_{i=1}^{N}\cosh^2\alpha_i) \sum_{l=1}^{N}\sinh^2\alpha_l \prod_{m=l+1}^{N}\cosh^2\alpha_m}{[1+(-1+\epsilon')\prod_{i=1}^{N} \cosh^2\alpha_i]^2}\nonumber\\
&-&\frac{\epsilon'^2\prod_{h=1}^{N} \cosh^4\alpha_h }{p[1+(-1+\epsilon')\prod_{i=1}^{N} \cosh^2\alpha_i]^2}\Big{]}\nonumber\\
&+&\zeta V\frac{\epsilon'^2 \prod_{h=1}^{N}\cosh^5\alpha_h}{[1+(-1+\epsilon')\prod_{i=1}^{N} \cosh^2\alpha_i]^2}\,.\nonumber
\enea
These are the main results of this article.

The transport coefficients can be rewritten in terms of the independent thermodynamic variables  of the reduced theory, the temperature and the chemical potentials, using Eq.\eqref{tempe} and Eq.\eqref{mu} functions of the rapidities.

Observe that the viscosity to entropy density ratio remains constant under the reduction,
\beq\label{KSS}
\frac{\hat\eta}{\hat s}=\frac{\eta}{s}\,.
\eneq

Furthermore, since the entropy current for our charged fluid in a canonical form is
\beq
\hat J_s^a=\hat s\hat u^a-\frac{u_b}{\hat T} T^{ab}_{diss}-\frac{1 }{\hat T}\sum_{j=1}^N \mu_j T^{ay_j}_{diss}
\eneq
using the relations in Eq.\eqref{eqn:Laundau1} and substituting the values of the chemical potentials Eq.(\ref{mu}), it is easy to see that
\beq
\hat J_s^a=\hat s\hat u^a\,.
\eneq
Comparing this result with the entropy density of our neutral initial fluid
\beq
\hat J_s^A=s u^A\,
\eneq 
multiplied by the volume factor $V$, we recover the result obtain from the Euler relation in Eq.(\ref{redentropy}).

Finally, the speed of sound is given by
\bea\label{sos}
\hat c_s^2=\frac{\partial\hat P}{\partial\hat\epsilon}=\frac{1}{ 1+(-1+\epsilon')\prod_{i=1}^{N} \cosh^2\alpha_i }\,,
\enea
 where the derivative is considered at fixed $\hat s/\hat q_j$ for every $ \hat q_j$.\\
%%%%%%%%%%%%%%%
\section{Charged black brane/fluid duals}
\label{6}

The previous analysis can be applied to the case of the fluid dual to
 a black p-brane. Let us consider a black p-brane in $D = p + n + 3$ dimensions with $p+1$ worldvolume coordinates of the p-brane and $n+2$ coordinates  in directions transverse
to that. Since we perform a Kaluza Klein reduction exclusively on the wordvolume directions, we focus only on the $p+1$ coordinates.  This means that the $p$ dimensions of the black p-brane can be seen as the $p$ spatial dimensions of the previous fluid.

In \cite{Joan} was shown that the long-wavelength dynamics of a neutral black brane in $D = p+ n + 3$ dimensions can be described in terms of a fluid with equation of state 
\beq\label{eqn:p}
\epsilon=-(n+1)P\,,
\eneq
and viscosities
\bea
\eta&=& \frac{s}{4\pi}\,,\\
\zeta&=&2 \eta\left(\frac{1}{p}+\frac{1}{n+1}\right)\label{eqn:etazeta}\,.
\enea

If we substitute these values in Eqs.\eqref{en}, \eqref{car}, \eqref{eqn:etaflui}, \eqref{eqn:kappafluidjj}, \eqref{eqn:kappafluidjk} and \eqref{eqn:zetafluid} we obtain the reduced thermodynamic quantities and the transport coefficients of the charged fluid. These are

\bea
\hat P&=&P  V, \qquad \hat \epsilon =-\hat P (1+n\prod_{i=1}^{N} \cosh^2\alpha_i )\,,\\
\hat q_j&=&-\hat P n \sinh\alpha_j \prod_{i=1}^{N} \cosh\alpha_i \prod_{k=1}^{j-1} \cosh\alpha_k\,,\\
\hat\eta&=&\eta V \prod_{i=1}^{N} \cosh\alpha_i\label{eqn:etargen}\\&=&\frac{\Omega_{n+1}V}{16 \pi G}\Big{(}\frac{ 4 \pi \hat T}{n}\Big{)}^{-n-1}(1-\sum_{i=1}^N \mu_i^2)^{\frac{n}{2}}\,,\nonumber\\
\enea
\bea
\hat\kappa_{jj}&=& \eta V\hat T \Big{(}1-\frac{\sinh^2\alpha_j}{\prod_{i=j}^{N} \cosh^2\alpha_i}\Big{)}\prod_{i=1}^{N} \cosh\alpha_i\label{eqn:coefkappagen}\\
&=&\frac{\Omega_{n+1}V}{16 \pi G}\Big{(}\frac{ 4 \pi }{n}\Big{)}^{-n-1}\hat T^{-n}(1-\mu_j^2)(1-\sum_{i=1}^N \mu_i^2)^{\frac{n}{2}},\nonumber\\
\hat\kappa_{jk}&=&-\eta V\hat T\frac{\sinh\alpha_j\sinh\alpha_{k}}{\prod_{i=j}^{N} \cosh\alpha_i}\prod_{l=1}^{k-1} \cosh\alpha_l\,\\&=&-\frac{\Omega_{n+1}V}{16 \pi G}\Big{(}\frac{ 4 \pi }{n}\Big{)}^{-n-1}\hat T^{-n}(\mu_k\mu_j)(1-\sum_{i=1}^N \mu_i^2)^{\frac{n}{2}},\nonumber\\
\hat\zeta&=&2\eta V\prod_{i=1}^{N} \cosh\alpha_i\Big{[} \frac{1}{p-N}\label{eqn:zetared}\\
&-&\frac{2\prod_{i=1}^{N}\cosh^2\alpha_i-(n+2)\prod_{i=1}^{N}\cosh^4\alpha_i - 1}{[1-(n+2)\prod_{i=1}^{N} \cosh^2\alpha_i]^2}\Big{]}\nonumber\\
&=&2\hat \eta\Big{\{} \frac{1}{p-N}- \frac{[-1-n-(\sum_{i=1}^N \mu_i^2)^2]}{[-1-n-\sum_{m=1}^N \mu_m^2]^2}\Big{\}},
\enea
using the explicit values of the temperature and the shear viscosity
\bea\label{identconst}
T=\frac{n}{4 \pi r_0}\,,\qquad\eta=\frac{\Omega_{n+1}}{16 \pi G}r_0^{n+1}
\enea
where $r_0$ is the horizon radius of the black p-brane and $\Omega_{n+1}$ is the volume of the unit (n+1)-sphere.

We can compare our results with those found in Eq.(3.4.17) and Eqs. (3.4.38)-(3.4.40) in \cite{Blaise} for $N=1$ in AdS. This can be done using the AdS/Ricci flat map in \cite{Joan2} which relates the dynamics of Ricci-flat black $p$-branes in $n+p+3$ dimensions to that of black $d$-branes in AdS$_{2\sigma+1}$, by identifying 
\beq\label{map}
-n\rightarrow 2\sigma\,,\qquad
 p\rightarrow d\,.\qquad
\eneq

If we apply this map to the equation of state and the transport coefficients that we obtain for $N=1$, we find the same results as in \cite{Blaise}
with
\bea
\alpha_i\rightarrow\omega_i\qquad \mbox{and}\qquad \frac{\Omega_{n+1}V}{16 \pi G}\rightarrow-L.
\enea
where $L$ is defined after Eq.(3.1.3) in [5].

For $N=2$ the map to black $d$-branes in AdS$_{2\sigma+1}$ is
\beq\label{map2}
-n\rightarrow 2\sigma\,,\qquad
 p\rightarrow d+1.
\eneq
In this case we recover the results of \cite{Milena} in Eq.(3.1.18) and Eqs. (3.2.42)-(3.2.51) for the Kaluza-Klein Einstein-Maxwell-Dilaton theory  containing two Maxwell fields, three neutral scalars and an axion in AdS, again using Eqs.\eqref{identconst}.
Note that the speed of sound and the other thermodynamic quantities as entropy, temperature and chemical potential agree too.

Finally, let us study whether the bound
\bea\label{bound}
\frac{\hat \zeta}{\hat \eta}\geq 2\left(\frac{1}{p-N}-\hat c^2_s\right)
\enea
 proposed in \cite{Buchel} is satisfied.
The bulk viscosity for a charged black $(p-N)$-brane takes the form
\bea
\hat\zeta&=&2\eta V\prod_{i=1}^{N} \cosh\alpha_i\Big{[} \frac{1}{p-N}\nonumber\\
&-&\frac{2\prod_{i=1}^{N}\cosh^2\alpha_i-(n+2)\prod_{m=1}^{N}\cosh^4\alpha_m - 1}{[1-(n+2)\prod_{i=1}^{N} \cosh^2\alpha_i]^2}\Big{]}\,.
\enea
In terms of the speed of sound
 \beq
\hat c_s^2=\frac{1}{1-(2+n)\prod_{i=1}^{N} \cosh^2\alpha_i }\,,
\eneq
the bulk to shear viscosity ratio can be written as
\bea
\frac{\hat \zeta}{\hat \eta}&=&2 \left(\frac{1}{p-N}-\hat c^2_s\right)\nonumber\\
&-&2\hat c^4_s \left[(n+4)\prod_{i=1}^{N}\cosh^2\alpha_i-(n+2)\prod_{m=1}^{N}\cosh^4\alpha_m -2\right].
\enea
The relation (\ref{bound}) requires that
\bea
n\geq -2+\frac{2}{\prod_{i=1}^{N}\cosh^2\alpha_i}\,,
\enea
which is satisfied for all $n\geq 0$. Thus the bound is always satisfied. In contrast, the bound is always violated in \cite{Blaise,Milena} for the black $d$-branes with $\sigma> 1$, where $2\sigma+1$ are the initial spacetime dimensions. 

An alternative bound was proposed in \cite{Blaise},
\bea\label{bound2}
\frac{\hat \zeta}{\hat \eta}\geq 2\left(\frac{1}{p-N}-\hat c^2_q\right)
\enea
in terms of the `speed of sound at constant charge density'
\bea
\hat c_q=\frac{\partial \hat P}{\partial \hat \epsilon}\Big{|}_{q_j}=\frac{2\prod_{l=1}^{N} \cosh^2\alpha_l-1}{ 1-(2+n)\prod_{i=1}^{N} \cosh^2\alpha_i }\,.
\enea
It is straightforward to show that in our case this bound is always violated. We obtain
\bea
\frac{\hat \zeta}{\hat \eta}&=&2 \left(\frac{1}{p-N}-\hat c^2_q\right)\nonumber\\
&-&2\hat c^4_s \left[(2+n)\prod_{i=1}^{N}\cosh^2\alpha_i(-1+\prod_{m=1}^{N}\cosh^2\alpha_m )\right]\,.
\enea
In order to satisfy the bound in (\ref{bound2}) we would need $n\le -2$.
The inversion of the results regarding both bounds \eqref{bound} and \eqref{bound2} as compared to  \cite{Blaise} and  \cite{Milena} is expected from the mappings \eqref{map} and \eqref{map2}. 
On the other hand,  for electrically charged, non-dilatonic asymptotically flat black
brane solutions,  ref.~\cite{Gath} finds that the bound \eqref{bound} is satisfied only for small enough charge density,
while the bound \eqref{bound2} is always violated.\footnote{Jakob Gath informs us (private communication) that for sufficiently large values of the dilaton coupling these bounds are satisfied/violated in the same manner as we have found: \eqref{bound} is always satisfied, and \eqref{bound2} is always violated, for all values of the charge density.}

Note Eq.(\ref{KSS}) implies that the KSS bound \cite{Kovtun} is saturated.

\section{Acknowledgements}

I am grateful to Roberto Emparan for discussions that led to this project, for many conversations, and  for carefully reading the draft. It is also a pleasure to thank Harvey Reall and Joan Camps for useful discussions and comments, and the warm hospitality provided by the Department of Applied Mathematics and Theoretical Physics of Cambridge University where part of this work was done.  
This work was supported by FPI scholarship BES-2011-045401 and grants MEC FPA2010-20807-C02-02, AGAUR 2009-SGR-168 and CPAN CSD2007-00042 Consolider-Ingenio 2010.

\appendix

%%%%%%%%%%%
\section{Equation of motion}\label{eqmotion}
In order to extract the transport coefficient we need to compute the equations of motion derived from energy momentum conservation relations.
These read as
\bea
\partial^a T^{pf}_{ab}&=&0=\partial^a[(\epsilon V+\hat P) \hat u_a \hat u_b \prod_{i=1}^{N} \cosh^2\alpha_i  + \hat P\eta_{ab}]\label{eqn:tuttogen}\\
&=&[ (1+\epsilon')\hat u_a\hat u_b \prod_{i=1}^{N} \cosh^2\alpha_i] \partial^a \hat P+\partial_b \hat P\nonumber\\
&+&(\epsilon V+\hat P ) \prod_{i=1}^{N} \cosh^2\alpha_i\Big[2\hat u_a\hat u_b\sum_{k=1}^{N} \tanh\alpha_k\partial^a \alpha_k
+\hat\theta\hat u_b+ \hat u_a\partial^a \hat u_b\Big]\,,\nonumber\\
\partial^a T^{pf}_{ay_j}&=&\partial^a(\epsilon V + \hat P)\sinh\alpha_j\hat u_a\prod_{i=1}^{N} \cosh\alpha_i \prod_{k=1}^{j-1} \cosh\alpha_k\label{eqn:loggen21}\\
&=&\{(1+\epsilon')\hat u_a \partial^a \hat P+(\epsilon V+\hat P )\Big[\sum_{i=1}^{N}\tanh\alpha_i
\hat u_a\partial^a\alpha_i \qquad\nonumber\\
&+&\sum_{l=1}^{j-1}\tanh\alpha_l
\hat u_a\partial^a\alpha_l+\coth\alpha_j\hat u_a\partial^a\alpha_j+\hat\theta\Big]\}\sinh\alpha_j\nonumber \\&&\prod_{i=1}^{N} \cosh\alpha_i \prod_{k=1}^{j-1} \cosh\alpha_k=0\nonumber\,
\enea
where
\beq\epsilon'=\frac{\partial \epsilon}{\partial P}\qquad\mbox{and}\qquad
\frac{\partial \epsilon}{\partial \hat P}=\epsilon' \frac{1}{V}\,.\nonumber
\eneq
If we contract the Eq.(\ref{eqn:tuttogen})  with $\hat u^b$  we find
\beq\label{eqn:loggen}
\hat u_a \partial^a \log \hat P=\frac{(\epsilon V + \hat P)} {\hat P}\prod_{i=1}^{N}\cosh^2\alpha_i\frac{-2\sum_{k=1}^{N} \tanh\alpha_k\hat u_a\partial^a \alpha_k-\hat\theta}{ (1+\epsilon')\prod_{i=1}^{N} \cosh^2\alpha_i- 1}\,.
\eneq
In addition, Eq.\eqref{eqn:loggen21} can be rewritten as
\bea
\hat u_a \partial^a \log \hat P=-\frac{(\epsilon V+\hat P )}{\hat P(1+\epsilon')}\Big[\sum_{i=1}^{N}\tanh\alpha_i
\hat u_a\partial^a\alpha_i \qquad\nonumber\\
+\sum_{l=1}^{j-1}\tanh\alpha_l
\hat u_a\partial^a\alpha_l+\coth\alpha_j\hat u_a\partial^a\alpha_j+\hat\theta\Big].\label{eqn:loggen2}
\enea
If we take the latter relation for two different indices, $j$ and $k$ with $j\neq k$ (corresponding to the conservation of two different components of the stress energy tensor) and we subtract them, we obtain
\beq\label{eqn:relazvarjdif}
\sum_ {i=k}^{j-1}\tanh \alpha_{i}\hat u_a \partial^a \alpha_i =\coth \alpha_{k}\hat u_a \partial^a \alpha_{k}-\coth \alpha_{j}\hat u_a \partial^a \alpha_{j}\,,\qquad j>k
\eneq
or equivalently
\beq
\hat u_a \partial^a \alpha_{i+1}=\hat u_a \partial^a \alpha_i \frac{\tanh{\alpha_{i+1}}}{\sinh{\alpha_{i}}\cosh{\alpha_{i}}}.
\eneq
Now, let us compare Eq.\eqref{eqn:loggen} and Eq.\eqref{eqn:loggen2}. This gives
\bea\label{sumconserv}
&&\prod_{i=1}^{N}\cosh^2\alpha_i\frac{-2\sum_{k=1}^{N} \tanh\alpha_k\hat u_a\partial^a \alpha_k-\hat\theta}{ (1+\epsilon')\prod_{i=1}^{N} \cosh^2\alpha_i- 1}=\\
&&-\frac{1} {(1+\epsilon')}
[\sum_{i=1}^{N} \tanh\alpha_i\hat u_a\partial^a \alpha_i+\sum_{l=1}^{j-1}\tanh\alpha_l\hat u_a\partial^a \alpha_l +\coth\alpha_j\hat u_a\partial^a \alpha_j+\hat\theta]\nonumber\\
\enea
Since we want the relation between the reduced expansion
and a derivatives of specific rapidity $\alpha_j$, we replace in Eq. \eqref{sumconserv} the other derivatives of the remaining rapidities in term of the one chosen
using Eq.\eqref{eqn:relazvarjdif}.
This gives  
\bea
\hat u_a \partial^a \alpha_j=\hat\theta \frac{\prod_{i=1}^{j-1}sech^2\alpha_i \tanh\alpha_j}{-1+\epsilon'+\prod_{l=1}^{N}sech^2\alpha_l}
\enea
that can be rewritten as
\beq\label{eqn:alfathetagen}
\hat u_a \partial^a \alpha_j =\hat\theta\frac{\cosh\alpha_{j}\sinh\alpha_{j}\prod_{l=j+1}^{N} \cosh^2\alpha_l}{1+(-1+\epsilon')\prod_{i=1}^{N} \cosh^2\alpha_i}.
\eneq

The Eq.(\ref{eqn:loggen}) in terms of $\hat u_a \partial^a \alpha_j$ only is given by  
\bea\label{eqn:lopPrapidgen}
\hat u_a \partial^a \log \hat P &=&-\frac{\epsilon V+\hat P}{\hat P}\coth \alpha_{j} \prod_{i=1}^{j-1}\cosh^2\alpha_{i}\hat u_a \partial^a \alpha_j
\enea
using the result in Eq.\eqref{eqn:alfathetagen}.

The reduced acceleration is obtain from Eq.\eqref{eqn:tuttogen} 
\bea
\hat u_a\partial^a \hat u_b&=&-\frac{\partial_b \log \hat P}{(1+ \frac{\epsilon V}{\hat P}) \prod_{i=1}^{N} \cosh^2\alpha_i}-\frac{\hat P (1+\epsilon')}{\hat P +\epsilon V}\hat u_b\hat u_a\partial^a \log \hat P \\
&-&2\hat u_a\hat u_b\sum_{k=1}^{N} \tanh\alpha_k\partial^a \alpha_k-\hat u_b\hat\theta\,.\nonumber
\enea
Using Eqs.\eqref{eqn:relazvarjdif}, \eqref{eqn:alfathetagen}, and \eqref{eqn:lopPrapidgen}, this becomes 
\beq\label{accel}
\hat u_a \partial^a\hat u_b=-\frac{\partial_b \log \hat P}{(1+ \frac{\epsilon V}{\hat P}) \prod_{i=1}^{N} \cosh^2\alpha_i}+ \frac{\hat u_a\hat u_b \partial^a \alpha_j}{\cosh \alpha_j\sinh \alpha_j \prod_{i=j+1}^{N}\cosh^2\alpha_{i}}\,.\eneq

%%%%%%%%%%%%%

\section{Transport coefficients}
\subsection{Shear viscosity}
The first term in Eq.(\ref{eqn:etagen}) is given by
\bea\label{firstter}
&&\hat P^a_c \hat P^b_d T_{ab}^{diss}=- V \hat P^a_c \hat P^b_d(2\eta\sigma_{ab}+\zeta P_{ab}\theta)\nonumber\\
&=&- V \hat P^a_c \hat P^b_d\Big[2\eta\Big( \sum_{i=1}^N P_{(a}^l  P^{y_i}_{b)}\partial_l u_{y_i}
+  P_a^l  P_b^m\partial_{(l}  u_{m)}- \frac{P_{ab}\theta}{p}\Big)+\zeta P_{ab}\theta\Big]\,.\qquad
\enea
Taking into account that
\bea
 \hat P_c^a\hat P_d^b  P_a^l P_b^m= \hat P_c^l\hat P_d^m \,,\quad\hat P_c^l \hat P_d^m\partial_{(l} \hat u_{m)}=0\,,\quad \hat P_c^a P_a^{yj}=0
\enea
the Eq.(\ref{firstter}) becomes
\bea\label{shear1}
\hat P^a_c \hat P^b_d T_{ab}^{diss}=- V\Big[2\eta \Big(\prod_{i=1}^N \cosh\alpha_i \hat P_c^l \hat  P_d^m\partial_{(l}  \hat u_{m)}- \frac{\hat P_{cd}\theta}{p}\Big)+\zeta \hat P_{cd}\theta\Big]\,.
\enea
For what concerns the second term in Eq.(\ref{eqn:etagen}) we get 
\bea
&& \frac{1}{p-N} \hat P_{cd} \hat P^{ab} T_{ab}^{diss}\qquad\label{shear2}\\
&=&-V\Big[2\eta \Big( \frac{1}{p-N}\prod_{i=1}^{N} \cosh\alpha_i \hat P_{cd}\hat P^{ab} \partial_{(a}\hat u_{b)}-
\frac{\hat P_{cd}}{p}\theta\Big)+\zeta \hat P_{cd}\theta\Big]\qquad\nonumber
\enea
considering that
\beq\label{abab}
\hat P^{ab} P_{ab}=p-N\,,\quad \hat P^{cd}\partial_{(c}\alpha_i \hat u_{d)}=0\,,\qquad\hat P^{ab} P^c_{(a}  P_{b)}^{y_i}= 0\,.
\eneq
If now we subtract (\ref{shear2}) with (\ref{shear1}) we are able to extract the shear viscosity. In fact
\bea
&-&2 V\eta \prod_{i=1}^N \cosh\alpha_i\Big[ \hat P_c^l \hat  P_d^m\partial_{(l}  \hat u_{m)}-\frac{1}{p-N} \hat P_{cd}\hat P^{ab} \partial_{(a}\hat u_{b)}\Big]\qquad\\
&=&-2 V\eta \prod_{i=1}^N \cosh\alpha_i\Big[ \hat P_c^l \hat  P_d^m\partial_{(l}  \hat u_{m)}- \frac{1}{p-N} \hat P_{cd}\hat\theta\Big]=-2\hat\eta\hat\sigma_{cd}\nonumber\,,
\enea
where $\hat P^{ab} \partial_{(a}\hat u_{b)}=\hat \theta$ and  $\hat\sigma_{cd}$ is defined as
\bea\label{eqn:sigmared}
\hat\sigma_{cd}= \hat P_c^a \hat P_d^b\partial_{(a} \hat u_{b)}-\frac{1}{p-N}\hat\theta \hat P_{cd}\,.
\enea
We recover the result anticipated in Eq.(\ref{eqn:etaflui}), that is
\bea
\hat\eta=\eta V \prod_{i=1}^{N} \cosh\alpha_i\,.
\enea

%%%%%%%%%%%%%%%%%%%%%
\subsection{Heat conductivity matrix}
Let us turn to the heat conductivity matrix elements. 
The Eq.(\ref{eqn:kappagen}) can be simplified substituting the Eqs.(\ref{eqn:Laundau1})  that leads to
\bea\label{eqn:kappagen1}
&&\hat P_a^b\Big( T_{by_j}^{diss}-\frac{\sinh\alpha_j}{\prod_{i=j}^{N} \cosh\alpha_i}\sum_{j'=1}^{N} \frac{\sinh\alpha_{j'}}{\prod_{i'=j'}^{N} \cosh\alpha_{i'}}  T_{by_{j'}}^{diss}\Big)\,
\enea
where we have also replaced the values of the reduced density charge, pressure and energy density from Eqs.(\ref{pres}), (\ref{en}) and (\ref{car}).
It is convenient to split  the above expression into two parts: one with index  $j'=j$ and the other with different indeces $k\neq j$ as
\bea
\hat P_a^b T_{by_j}^{diss}\Big(1-\frac{\sinh^2\alpha_j}{\prod_{i=j}^{N} \cosh\alpha^2_i}\Big)-\hat P_a^b\frac{\sinh\alpha_j}{\prod_{i=j}^{N} \cosh\alpha_i}\sum_{k=1}^{N} \frac{\sinh\alpha_{k}}{\prod_{i'=k}^{N} \cosh\alpha_{i'}}  T_{by_{k}}^{diss}.\nonumber
\enea
 So, we extract the heat conductivity coefficients from the relations
\bea
\hat P_a^b T_{by_j}^{diss}\Big(1-\frac{\sinh^2\alpha_j}{\prod_{i=j}^{N} \cosh\alpha^2_i}\Big)&=&-\hat\kappa_{jj} \hat P^b_a\partial_b\left(\frac{\hat\mu_{j}}{\hat T}\right)\label{kappajj}\,,\qquad\\
-\hat P_a^b\frac{\sinh\alpha_j}{\prod_{i=j}^{N} \cosh\alpha_i}\sum_{k=1}^{N} \frac{\sinh\alpha_{k}}{\prod_{i'=k}^{N} \cosh\alpha_{i'}}  T_{by_{k}}^{diss}&=&-\sum_{k=1}^{N}\hat\kappa_{jk} \hat P^b_a\partial_b\left(\frac{\hat\mu_{k}}{\hat T}\right)\,.\qquad\label{kappajk}
\enea
The $\hat P_a^b T_{by_j}^{diss}$  term is given by
\bea
\hat P_a^b T_{by_j}^{diss}&=&-2\eta V \hat P_a^{c}\Big(P^d_{y_j}\partial_{(c} u_{d)}+\frac{1}{2}\sum_{j'=1}^{N} P_{y_j}^{y_{j'}}\partial_c u_{y_{j'}}\Big)\\
&=&-\eta V \hat P_a^{c}\Big[\sinh\alpha_j \prod_{l=1}^{j-1}\cosh\alpha_l\prod_{l'=1}^{N}\cosh\alpha_{l'}\label{PabTby}\nonumber\\
&&(-\partial_c \prod_{i'=1}^{N}\cosh\alpha_{i'}+\prod_{i=1}^{N}\cosh\alpha_{i}\hat u_b\partial^b\hat u_c )+\sum_{j'=1}^{N} P_{y_j}^{y_{j'}}\partial_c u_{y_{j'}}\Big]\,.\nonumber
\enea
Using the expression (\ref{accel}) for the acceleration, and the fact that $\hat P^{ab}u_b=0$, this becomes
\bea
&-&\eta V \hat P_a^{c}\Big[\sinh\alpha_j \prod_{l=1}^{j-1}\cosh\alpha_l\prod_{l'=1}^{N}\cosh\alpha_{l'}\nonumber\\
&&\Big(-\partial_c \prod_{i'=1}^{N}\cosh\alpha_{i'} -\frac{\partial_c \log \hat P}{(1+\frac{\epsilon V}{\hat P})\prod_{i=1}^{N}\cosh\alpha_{i}}\Big)
+\sum_{j'=1}^{N} P_{y_j}^{y_{j'}}\partial_c u_{y_{j'}}\Big]\nonumber
\enea
Making explicit the value of $P_{y_j}^{y_i}$ and simplifying we obtain
\bea
&-&\eta V \hat P_a^{c}\sinh\alpha_j \prod_{l=1}^{j-1}\cosh\alpha_l\Big(\sum_{l'=1}^{j-1}\tanh\alpha_{l'}\partial_c \alpha_{l'}+ \coth\alpha_j\partial_c \alpha_{j}\nonumber\\
&-&\frac{\partial_c \log \hat P}{1+\frac{\epsilon V}{\hat P}}\Big)\label{PabTbyfin}\,,
\enea

For what concerns the right part of the formula (\ref{kappajj}), if we substitute the values (\ref{tempe}) and (\ref{mu}) we find
\bea
&&\hat P^b_a \partial_b \left(\frac{\hat\mu_j}{\hat T}\right)= \hat P^b_a \partial_b \Big(\frac{\sinh\alpha_j\prod_{i=1}^{j-1}\cosh\alpha_i}{T}\Big)\label{derivmuT}\\
&&=\frac{\hat P^b_a}{\hat T} \frac{\sinh\alpha_j}{\prod_{i=j}^{N}\cosh\alpha_i}\Big[ \coth\alpha_j \partial_b\alpha_j+\sum_{l=1}^{j-1} \tanh\alpha_{l}\partial_b \alpha_l-\frac{\partial_b\log \hat P}{1+\frac{\epsilon V}{\hat P}}\Big]\label{eqn:condu2gen}\,.\qquad
\enea

Now we are able to extract the elements of the heat conductivity metric from Eq.(\ref{kappajj}) only replacing the result obtained in Eq.(\ref{PabTbyfin}) and (\ref{eqn:condu2gen}). This leads to
\bea
&-&\eta V \hat P_a^{c}\sinh\alpha_j \prod_{l=1}^{j-1}\cosh\alpha_l\biggl[\sum_{l'=1}^{j-1}\tanh\alpha_{l'}\partial_c \alpha_{l'}+ \coth\alpha_j\partial_c \alpha_{j}\\
&-&\frac{\partial_c \log \hat P}{(1+\frac{\epsilon V}{\hat P})}\biggr]\Big(1-\frac{\sinh^2\alpha_j}{\prod_{i=j}^{N} \cosh\alpha^2_i}\Big)\nonumber\\
&=&-\hat\kappa_{jj}\frac{\hat P^c_a}{\hat T} \frac{\sinh\alpha_j}{\prod_{i=j}^{N}\cosh\alpha_i}\Big[ \coth\alpha_j \partial_c\alpha_j
+\sum_{l=1}^{j-1} \tanh\alpha_{l}\partial_c \alpha_l\\&-&
\frac{\partial_c \log \hat P}{1+\frac{\epsilon V}{\hat P}}]\,.\nonumber
\enea
So, the diagonal elements are
\bea
\hat\kappa_{jj}= \eta V\hat T\prod_{i=1}^{N} \cosh\alpha_i\Big(1-\frac{\sinh^2\alpha_j}{\prod_{i=j}^{N} \cosh\alpha^2_i}\Big)\,,
\enea
 as already shown in Eq.(\ref{eqn:kappafluidjj}).

The same procedure is performed for the Eq.(\ref{kappajk}).  We find that
\bea
&-&\hat P_a^b T_{by_k}^{diss}\frac{\sinh\alpha_j}{\prod_{i=j}^{N} \cosh\alpha_i}\sum_{k=1}^{N} \frac{\sinh\alpha_{k}}{\prod_{i'=k}^{N} \cosh\alpha_{i'}}\qquad\\
&=&\eta V\hat P_a^{c}\frac{\sinh\alpha_j}{\prod_{i=j}^{N} \cosh\alpha_i}\sum_{k=1}^{N} \frac{\sinh\alpha_{k}}{\prod_{i'=k}^{N} \cosh\alpha_{i'}}\Big[\sinh\alpha_k \prod_{l=1}^{k-1}\cosh\alpha_l\nonumber\\
&&\Big(\sum_{l'=1}^{k-1}\tanh\alpha_{l'}\partial_c \alpha_{l'}
+ \coth\alpha_k\partial_c \alpha_{k}
-\frac{\partial_c \log \hat P}{1+\frac{\epsilon V}{\hat P}}\Big)\Big]\nonumber\\
&=&-\sum_{k=1}^{N}\hat\kappa_{jk}\frac{\hat P^c_a}{\hat T} \frac{\sinh\alpha_k}{\prod_{i=k}^{N}\cosh\alpha_i}\Big[ \coth\alpha_k \partial_c\alpha_k
+\sum_{l=1}^{k-1} \tanh\alpha_{l}\partial_c \alpha_l-\frac{\partial_c \log \hat P}{1+\frac{\epsilon V}{\hat P}}\Big]\nonumber\,.
\enea
Comparing the last two equations, term by term, in the sum it is easy to find that
\bea
\hat\kappa_{jk}=-\eta V\hat T\frac{\sinh\alpha_j\sinh\alpha_{k}}{\prod_{i=j}^{N} \cosh\alpha_i}\prod_{l=1}^{k-1} \cosh\alpha_l\,.
\enea
%%%%%%%%%%

%%%%%%%%%%%%%%
\subsection{Bulk viscosity}
Let complete the hydrodynamic analysis computing the bulk viscosity. \\

First of all, we need to compute the derivatives of the pressure with respect to energy density and the charges. 
Due to the fact that $\partial\hat P/\partial\hat\epsilon$ is calculated with constant charges $\hat q_j$ and $\partial\hat P/\partial\hat q_j$ keeping fixed the energy density and the remaining $q_{k\neq j}$ charges, we need to use 
\bea
d \hat \epsilon=0 &\Rightarrow& d \log \hat P=-2\frac{ (\hat P+ \epsilon V) }{\hat P}\frac{\prod_{m=1}^{N} \cosh^2\alpha_m \sum_{l=1}^N\tanh\alpha_l}{[-1+(\epsilon'+1)\prod_{k=1}^{N}  \cosh^2\alpha_k]}d\alpha_l\nonumber\\\label{enetgconst}\\
d \hat q_j=0 &\Rightarrow& d \log \hat P=-\frac{ (\hat P+ \epsilon V) }{\hat P(1+\epsilon')}\Big[\sum_{l=1}^N\tanh\alpha_l d\alpha_l+ \coth\alpha_j d\alpha_j\nonumber\\
&+& \sum_{k=1}^{j-1}\tanh\alpha_k d\alpha_k\Big].\label{carconst}
\enea

The considered derivatives are given by
\bea\label{derivatives3}
&&\frac{\partial\hat P}{\partial\hat\epsilon}=\\
&&\frac{1 }{[-1+(1+\epsilon')\prod_{l=1}^{N} \cosh^2\alpha_l]+ 2\frac{\hat P+ \epsilon V}{\hat P}\prod_{i=1}^{N}\cosh\alpha_i^2\sum_{l=1}^N\tanh\alpha_l\frac{\partial \alpha_l}{\partial\log\hat P}}\nonumber,\\
&&\frac{\partial\hat P}{\partial\hat q_j}=\frac{1 }{(1+\epsilon')A+ \frac{\hat P+ \epsilon V}{\hat P}A B}\,,\label{derivatives4}
\enea
with 
\bea
A&=&\sinh\alpha_j\prod_{i=1}^{N} \cosh\alpha_i \prod_{k=1}^{j-1} \cosh\alpha_k \\
B&=&\sum_{l=1}^N\tanh\alpha_l\frac{\partial \alpha_l}{\partial\log\hat P}+\sum_{k=1}^{j-1}\tanh\alpha_k\frac{\partial \alpha_k}{\partial\log\hat P}+\coth\alpha_j\frac{\partial \alpha_j}{\partial\log\hat P}\nonumber\\
\enea

Combining conveniently Eqs.\eqref{enetgconst} and \eqref{carconst} in Eqs.\eqref{derivatives3} and \eqref{derivatives4} we obtain that

\bea\label{derivatives}
\frac{\partial\hat P}{\partial\hat\epsilon}&=&\frac{2\prod_{i=1}^{N}\cosh^2\alpha_i-1}{ 1+(-1+\epsilon')\prod_{l=1}^{N} \cosh^2\alpha_l},\label{derivatives1}\\
\frac{\partial\hat P}{\partial\hat q_j}&=&\frac{-2\sinh\alpha_j\prod_{i=1}^{N}\cosh\alpha_i\prod_{m=1}^{j-1}
\cosh\alpha_m}{ 1+(-1+\epsilon')\prod_{l=1}^{N} \cosh^2\alpha_l}\,.\label{derivatives2}
\enea

Let now evaluate the first term in Eq.(\ref{eqn:zetagen}).
We find that 
\bea\label{firstterm}
&& \frac{\hat P^{ab} T_{ab}^{diss}}{p-N}\qquad\\
&&=-\hat \theta V \prod_{l=1}^{N} \cosh\alpha_l\left[ \frac{2\eta}{p-N}
+\left(\frac{-2\eta}{p}+\zeta\right) \frac{\epsilon'\prod_{m=1}^{N} \cosh^2\alpha_m}{1+(-1+\epsilon')\prod_{i=1}^{N} \cosh^2\alpha_i}\right]\,.\nonumber
\enea

The remaining terms can be simplified considering the Eq.(\ref{eqn:Laundau1}) and Eq.(\ref{eqn:Laundau2}). These become
\bea
-\sum_{j'=1}^{N} \frac{\sinh\alpha_{j'}}{\prod_{m=j'}^{N}\cosh\alpha_m}\left(\frac{\partial \hat P}{\partial \hat\epsilon}\sum_{j=1}^{N} \frac{\sinh\alpha_j}{\prod_{i=j}^{N}\cosh\alpha_i}+\sum_{j=1}^{N}\frac{\partial \hat P}{\partial \hat q_j}\right)T_{y_jy_{j'}}^{diss}
\enea
Using the Eqs.(\ref{eqn:alfathetagen}), (\ref{eqn:relazvarjdif}), \eqref{derivatives}, \eqref{derivatives2} and replacing the components of the stress energy tensor $T_{y_jy_{j'}}^{diss}$ we find 
\bea
&&-\hat\zeta \hat \theta=-2\eta\hat \theta V\prod_{i=1}^{N} \cosh\alpha_i\Big{[} \frac{1}{p-N}\qquad\\
&&+\frac{(-1+\prod_{i=1}^{N}\cosh^2\alpha_i) \sum_{l=1}^{N}\sinh\alpha_l^2 \prod_{m=l+1}^{N}\cosh^2\alpha_m}{[1+(-1+\epsilon')\prod_{i=1}^{N} \cosh^2\alpha_i]^2}\nonumber\\
&&-\frac{\epsilon'^2\prod_{h=1}^{N} \cosh^4\alpha_h }{p[1+(-1+\epsilon')\prod_{i=1}^{N} \cosh^2\alpha_i]^2}\Big{]}-\zeta\hat \theta V\frac{\epsilon'^2 \prod_{h=1}^{N}\cosh^5\alpha_h}{[1+(-1+\epsilon')\prod_{i=1}^{N} \cosh^2\alpha_i]^2}\nonumber\,.
\enea
in terms of $\hat\theta$ only. Remember that for the initial expansion we use the values found in the Eq.(\ref{eqn:thetagen1}) .
It is straightforward to read now the bulk viscosity as already presented in Eq.(\ref{eqn:zetafluid}).

\end{document}